\documentclass{aa}  

\usepackage{graphicx}
\usepackage{txfonts}
\usepackage{lipsum}
\usepackage{subcaption}
\usepackage{lscape}    
\usepackage{float}
                               
\usepackage{placeins}           

\usepackage{hyperref}

\begin{document}

   \title{Binary stars as dynamical tracers in globular clusters .II. }\subtitle{Evolution of the radial distribution of the binary star fraction}

   \author{J. Bruce\inst{1}\fnmsep\thanks{E-mail: jobruce@iu.edu}
        \and E. Vesperini \inst{1}
        \and E. Dalessandro \inst{2}
        \and M. Cadelano \inst{2,3}
        \and F. R. Ferraro \inst{2,3}
        \and B. Lanzoni \inst{2,3}
        \and G. Beccari \inst{4}
        \and C. Giusti \inst{2, 3}
        }

   \institute{Department of Astronomy, Indiana University, Swain West, 727 E. 3rd Street, Bloomington, IN 47405, USA
   \and 
   INAF – Astrophysics and Space Science Observatory of Bologna, via Gobetti 93/3, I-40129 Bologna, Italy
   \and
   Department of Physics and Astronomy ‘Augusto Righi’, University of Bologna, via Gobetti 93/2, I-40129 Bologna, Italy
   \and
   European Southern Observatory, Karl-Schwarzschild-Strasse 2, 85748 Garching bei München, Germany}

   \date{Submitted --}
 
  \abstract
    {The distinct dynamical environments occupied by multiple stellar populations in globular clusters play a significant role in many aspects of cluster evolution, including the dynamics of binary stars. Recent observational analyses of Galactic globular clusters have revealed a bimodal radial trend in the binary fraction: the fraction of binaries is enhanced in both the central and outer regions of clusters, with a noticeable minimum in the cluster's intermediate regions. In this paper, we investigate the dynamical origin of this feature and the mechanisms responsible for preserving this bimodality for an extended portion of a cluster's lifetime. We utilize a suite of Monte Carlo simulations that follow the long-term dynamical evolution of both single-population and multiple-population globular clusters. We investigate how mass segregation and binary disruption function cohesively to produce a bimodal profile, and show that although single-population clusters can briefly generate a similar profile, the effect is weak and transient. Conversely, we show that the structural properties associated with the presence of multiple stellar populations significantly strengthen and preserve the bimodality. We also show that the effects of long-term dynamical evolution drive a broad relationship between a cluster's dynamical age and the radial location of the binary fraction minimum, which tends to migrate outward over time. Overall, our results strongly indicate that the typical multi-scale structure of multiple-population globular clusters (initially characterized by a dense and centrally concentrated subsystem of second-population stars embedded in a more extended first-population system) plays a key role in the origin of the observed bimodal profiles and they further demonstrate the analytical power of binary stars as dynamical tracers of globular cluster formation and dynamical evolution.}


   \maketitle
   \nolinenumbers

\section{Introduction}
\label{sec:1}
Binary stars are powerful tracers of the internal dynamics of star clusters and have contributed significantly to our understanding of the dynamical evolution of Galactic globular clusters (GCs). In dense stellar systems, binaries experience frequent dynamical encounters that can result in their disruption, the hardening or softening of their configurations, or their ejection from the system. The outcome of each interaction depends on the properties of the binary and on the local dynamical environment within the cluster. As a result, both the survival rate and spatial distribution of binaries carry valuable information regarding a cluster's initial configuration and dynamical history. 

The complex composition of GCs has been enlightened by observational studies showing the presence of multiple populations uniquely identified with distinct chemical compositions, in particular, a spread in various light elements such as Na, O, Al, Mg, C, and N (see, e.g., \citeauthor{Carretta+2009a} \citeyear{Carretta+2009a}a,b; \citeauthor{Milone+2018b} \citeyear{Milone+2018b}; \citeauthor{Marino+2019} \citeyear{Marino+2019}; \citeauthor{Carretta2019} \citeyear{Carretta2019}; and \citeauthor{Milone+2022} \citeyear{Milone+2022} for a recent review). The origin of multiple populations is still a matter of intense investigation but a common feature of many proposed formation scenarios is that the second stellar population (P2) forms more centrally concentrated than the first stellar population (P1), with P2 embedded within the inner regions of P1 (see, e.g., \citeauthor{Dercole+2008} \citeyear{Dercole+2008}, \citeauthor{Bekki2010} \citeyear{Bekki2010}, \citeauthor{Bastian+2013} \citeyear{Bastian+2013}, \citeauthor{Calura+2019} \citeyear{Calura+2019}, \citeauthor{Gieles+2018} \citeyear{Gieles+2018}; \citeyear{Gieles+2025}). Although two-body relaxation gradually reduces the initial differences between the stellar populations, the present-day properties of many GCs still show the fingerprints of those initial differences. In particular, a number of observational studies have revealed differences between the P1 and P2 kinematics (see, e.g., \citeauthor{Bellini+2015} \citeyear{Bellini+2015}; \citeauthor{Dalessandro+2018b} \citeyear{Dalessandro+2018b}, \citeyear{Dalessandro+2021}, \citeyear{Dalessandro+2024}; \citeauthor{Milone+2018a} \citeyear{Milone+2018a}; \citeauthor{Martens+2023} \citeyear{Martens+2023}; \citeauthor{Cadelano+2024} \citeyear{Cadelano+2024}; and \citeauthor{Cordoni+2025dynamics} \citeyear{Cordoni+2025dynamics}), and spatial distributions (see, e.g., \citeauthor{Bellini+2009} \citeyear{Bellini+2009}; \citeauthor{simioni+2016} \citeyear{simioni+2016}; \citeauthor{dalessandro+2016} \citeyear{dalessandro+2016}, \citeyear{Dalessandro+2018a}, \citeyear{Dalessandro+2019}; and \citeauthor{Onorato+2023} \citeyear{Onorato+2023}). 

The differences between the initial spatial distributions of the P1 and P2 populations also have significant implications for the evolution of their binary stars. The compact P2 subsystem is characterized by shorter relaxation times and higher encounter rates, enhancing the rate of binary disruption and quickening mass segregation. Alternatively, the more extended P1 population occupies less hostile regions where the velocity dispersion and encounter rates are lower, preserving a larger fraction of wide binaries in the cluster outskirts. These unique dynamical environments have major impacts on the radial distributions, the relative binary fractions in each population, and the long-term dynamical evolution of binary populations. These effects have been investigated in a number of numerical studies (see, e.g., \citeauthor{Vesperini+2011} \citeyear{Vesperini+2011}; \citeauthor{Hong+2015} \citeyear{Hong+2015}, \citeyear{Hong+2016}, \citeyear{Hong+2019}; \citeauthor{Hypki+2022} \citeyear{Hypki+2022}; and \citeauthor{Bruce+2026} \citeyear{Bruce+2026}), and a few observational studies (see e.g. \citeauthor{Lucatello+2015} \citeyear{Lucatello+2015}; \citeauthor{Dalessandro+2018b} \citeyear{Dalessandro+2018b}; \citeauthor{Kamann+2020} \citeyear{Kamann+2020}; \citeauthor{Milone+2020} \citeyear{Milone+2020}, \citeyear{Milone+2025}; \citeauthor{Bortolan+2025} \citeyear{Bortolan+2025}) have started to reveal differences between the binary incidence of the P1 and P2 populations consistent with theoretical predictions.

A recent study by \citet{Cadelano+2026PREP} (hereafter referred to as paper I) has revealed a striking feature in the radial distribution of binaries in GCs: a bimodal profile characterized by a peak in the central regions, a minimum at intermediate radii, and an additional rise in the outskirts (see also \citeauthor{Beccari+2013} \citeyear{Beccari+2013} for an early study hinting at a similar radial profile in NGC 5466, and \citeauthor{Cordoni+2025binary} \citeyear{Cordoni+2025binary} for an increased binary fraction in the outskirts of 47 Tuc compared to its central regions). Here, we briefly outline the main findings of paper I. The authors provide the first homogeneous investigation of the radial dependence of the binary fraction across six GCs using a combination of HST and ground-based observations. This sample includes GCs at different evolutionary stages, from dynamically young clusters to post-core-collapse clusters, and five of the six clusters display a bimodal binary fraction radial distribution (see their Fig. 5). Additionally, the observations highlight a trend between the location of the binary fraction minimum and the dynamical age of the cluster, where the location of the minimum is found at larger clustercentric radii for dynamically older GCs (see their Fig. 6). Finally, although extremely dynamically evolved systems (post core-collapse) are expected to display a monotonically decreasing binary fraction with increasing radius, one of the observed post-core-collapse clusters displays a bimodality, possibly revealing the effect of the scattering of binaries from the core to the outer regions of the cluster.

Earlier theoretical work \citep{Geller+2013} has shown that a similar bimodal profile can naturally arise in single population clusters as a result of binary disruption and mass segregation. However, in those simulations the bimodality was transient and significantly weaker compared to observations, which show an increase in binary fraction by a factor of 2 in the outer regions compared to the minimum in most clusters (see Fig. 5 of paper I). The presence of multiple stellar populations can, however, strengthen the bimodality and allow it to survive over much longer timescales.

In this paper, we investigate the long-term dynamical evolution of binary stars in GCs containing multiple stellar populations, with the goal of explaining the origin and evolution of the bimodal binary fraction profile. Using a series of Monte Carlo simulations, we investigate how binary disruption, mass segregation, and the typical structural properties of multiple-population clusters described above can produce and preserve the observed bimodality. Additionally, we present a broad relationship between the radial position of the binary fraction minimum and the clusters' dynamical ages.

The paper is organized as follows. In section \ref{sec:2}, we describe the simulations and analytical techniques. Section \ref{sec:3} presents our results, first for single population clusters (Section \ref{sec:3.1}) and then for systems with multiple populations (Section \ref{sec:3.2}), where we discuss the role of disruption and segregation. Additionally, we discuss the correlation between the radial location of the minimum of the binary fraction profile, the intersection of the P1 and P2 binary fractions, and the dynamical age of the clusters (Section \ref{sec:3.3}). Finally, Section \ref{sec:4} summarizes our main conclusions and discusses the implications of using binaries as dynamical tracers in globular clusters. 

\section{Methods}
\label{sec:2}

To investigate the dynamical origin of the bimodal radial profiles, we analyzed a set of Monte Carlo star cluster simulations run with the MOCCA code (see e.g. \citeauthor{Giersz1998} \citeyear{Giersz1998}, \citeauthor{Hypki+2013} \citeyear{Hypki+2013}). Our suite includes both single population (SP) and multiple population (MP) models and we follow their binary populations over a total evolution time of 12 gigayears.

We follow a similar methodology to \citet{Bruce+2026} which we briefly outline here and refer to that paper for further details. Each simulation starts with $10^6$ stars whose masses are drawn from a \citet{Kroupa2001} initial mass function between 0.1 and 100 $M_\odot$. Very few or no black holes are retained in our simulations; the retention of a significant fraction of black holes can affect the structural evolution of globular clusters toward core collapse, the degree of mass segregation and energy equipartition \citep[see e.g. ][]{Breen+2013, Kremer+2018, Askar+2018, Aros+2021, Aros+2023, dellacroce+2024}. We will address the investigation of the evolution of models similar to those studied here for different black hole retention fractions in a future investigation. A few recent studies \citep[see e.g.][]{Cadelano+2020, Baumgardt+2023} have suggested that globular clusters might form with an IMF bottom-light compared to the \citet{Kroupa2001} IMF. In future works we will consider the dynamics of multiple-population clusters with a bottom-light IMF (Bruce et al. in prep) but we do not expect a large effect on the specific results presented in this paper.

The initial binary fraction is set to 10$\%$, allowing the simulations to reach binary fractions comparable to many observed Galactic globular clusters after 12 Gyr. The SP model is initialized with a \citet{King1966} profile with central potential parameter $W_0$ = 7. In the MP simulations, we include two distinct stellar populations, with P1 and P2 following King models with $W_0$ = 5 and 7, respectively. We characterize the central concentration of P2 using the ratio of its half-mass radius to that of P1, which is set to 0.05 in these simulations. All models have an initial tidal radius ($r_t$) equal to about 59 pc (except the model evolving in the strong tidal field for which $r_t$ is equal to about 38 pc) and a ratio of the half-mass radius ($r_h$) to the tidal radius ($r_h/r_t$) equal to about 0.14 (except for the single-population cluster which has $r_h/r_t$ equal to about 0.017). We include in Appendix~\ref{sec:A1} the cumulative mass profile of the MP and the SP models showing they have a similar profile in the inner regions but the MP system is characterized by a multi-scale structure with a more extended low-density outer region (initially populated by P1 stars). In order to provide some indication of the dependence of our results on other parameters, we show in  Appendix~\ref{sec:B1} the radial profile of the binary fraction for two additional models, one with a binary fraction of $5\%$ and one with a ratio of P2 to P1 half-mass radii set to 0.1. This configuration places the more centrally concentrated P2 subsystem within the inner regions of P1, consistent with a number of scenarios and simulations regarding the formation of multiple populations (see, e.g., \citeauthor{Dercole+2008} \citeyear{Dercole+2008}, \citeauthor{Bekki2010} \citeyear{Bekki2010}, and \citeauthor{Calura+2019} \citeyear{Calura+2019}).

Binary semimajor axes are sampled uniformly in logarithmic space up to a maximum of 100 AU in both the SP and MP models. To examine the impact of wider binaries on the radial profiles, we also run a model in which the semimajor axis upper limit is increased to 1000 AU (MPwide model). In addition, we include a model (MPsf) in which the cluster evolves in a stronger tidal field. This model has a shorter relaxation time, experiences more extensive mass loss, and reaches older dynamical ages over the course of the 12 Gyr.

Some properties of the binary population in the MP and MPsf models were discussed in \citet{Bruce+2026}, while the SP and the MPwide models are introduced here for the first time.

To compare directly with the observational analyses of paper I, we focus on main sequence binaries with primary masses between 0.5 and 0.75 $M_\odot$ and mass ratios q $>$ 0.4, together with single main sequence stars in the same mass range. For each snapshot, we construct the projected radial profiles using 100 independent random 2D projections, and we report the median with shaded regions representing the 25th and 75th percentiles. We define the binary fraction for P1 as

\begin{equation}
    f_b^{P1} = \frac{N_b^{P1}}{N_b^{Tot} + N_s^{Tot}}
\end{equation}

\noindent the fraction of all observable objects from any population ($N_b^{Tot} + N_s^{Tot}$) in the cluster that are in a binary system from P1 ($N_b^{P1}$), and an analogous definition holds for P2.

\section{Results}
\label{sec:3}
\subsection{Single population}
\label{sec:3.1}

\begin{figure}[h]
    \centering
    \includegraphics[width=0.95\linewidth, height=0.65\textheight]{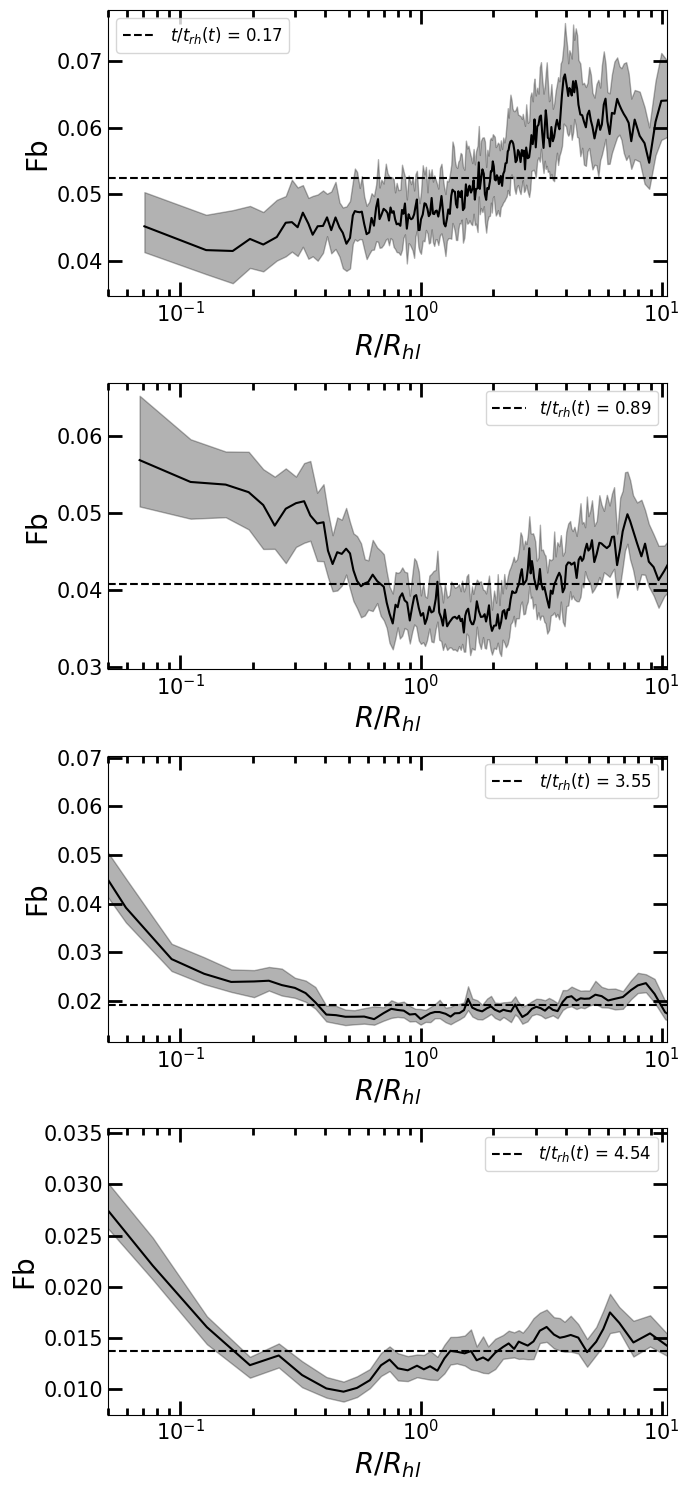}
    \caption{Radial profile of the fraction of binaries as a function of the projected distance from the cluster center normalized to the projected half-light radius for the SP simulation at four different times. We report the median of 100 random realizations of the 2D spatial projection, with the shaded regions representing the 25th and 75th percentiles. The horizontal dashed line displays the global value of the binary fraction.}
    \label{fig:fig 1}
\end{figure}

\begin{figure}[h!]
    \centering
    \includegraphics[width=0.95\linewidth, height=0.65\textheight]{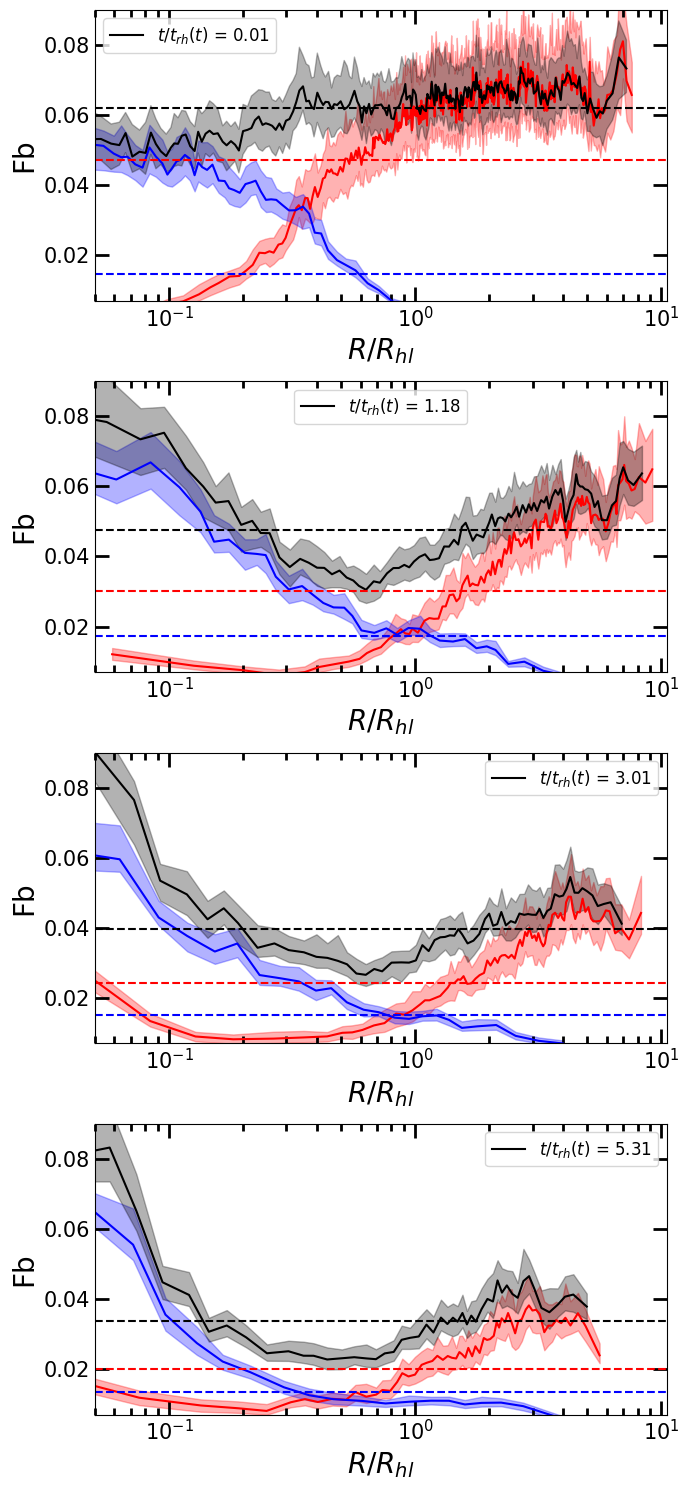}
    \caption{Radial profile of the fraction of binaries in P1 (red), P2 (blue), and both populations (black) as a function of the projected distance from the cluster center normalized to the projected half-light radius for the MP simulation at four different times. We report the median of 100 random realizations of the 2D spatial projection, with the shaded regions representing the 25th and 75th percentiles. The horizontal dashed lines display the global values of the binary fraction for P1 (red dashed line), P2 (blue dashed line) and P1+P2 (black dashed line).}
    \label{fig:fig 2}
\end{figure}

The evolution of the binary fraction radial profile for the SP simulation at four times is shown in Figure \ref{fig:fig 1}. We select four snapshots for the SP simulation and the MP simulations that correspond to similar dynamical ages within each model which were previously introduced (see Sect. \ref{sec:2}). We chose these times to show the dynamical evolution of the clusters at early, intermediate, and late stages. At early times, the combined effects of soft binary disruption and the survival of wide binaries in the outskirts lead to a monotonic outward increase in the binary fraction. This behavior is expected, as the dense central regions are characterized by a larger local velocity dispersion causing a larger fraction of binaries to fall in the soft regime, and the higher encounter rate further accelerates their disruption. Alternatively, the smaller velocity dispersions in the outer regions imply that wide binaries are less soft, and the fewer interactions make them more likely to survive. This is consistent with the trends presented in \citet{Geller+2013}, who reported that soft binary disruption naturally produces a monotonic profile, although only as a short lived feature.

As the cluster evolves, mass segregation becomes increasingly important. Hard binaries that can survive in the dense inner regions sink toward the center of the cluster, generating a central peak in the binary fraction. This process transforms the initially monotonic profile to become bimodal within the first 1-2 Gyr of the cluster's evolution. This result is once again consistent with \citet{Geller+2013}, however, we note that the bimodality reported in \citet{Geller+2013} is significantly less pronounced compared to both the observations and the results of our simulations. Despite the bimodality being naturally produced, it is highly transient, and binaries migrating from larger radii replenish the deficit at intermediate radii, efficiently restoring a more monotonic profile.

At later times, once the cluster has undergone core collapse, a mild secondary bimodality reappears (shown in the final panel of Figure \ref{fig:fig 1}). At this stage of cluster evolution, repeated encounters involving hard binaries in the core can impart recoil velocities (see, e.g., \citeauthor{HeggieHut2003} \citeyear{HeggieHut2003}) that scatter binaries to larger radii. This outward scattering slightly increases the binary fraction in the outskirts, producing a mild secondary increase. Nevertheless, this secondary bimodality is significantly weaker than that identified in the observations, the minimum lies closer to the cluster center, and the feature appears only 1-2 Gyr after core collapse (see Fig. 5 and 6 of paper I for observational comparison). These results suggest that the internal dynamics of single population clusters alone cannot account for the strong bimodality detected in observations, but the scattering of binaries post core collapse could result in a secondary rise in the binary fractions in the outer regions. Understanding whether this mechanism is responsible for producing the secondary bimodality warrants further investigation in a future study.

\subsection{Multiple population}
\label{sec:3.2}

A more defined and stable bimodality emerges in the MP simulation, displayed in Figure \ref{fig:fig 2}. Although the same dynamical processes are in effect, their impact on the radial profile of the total binary fraction is amplified by the presence and environmental differences of the multiple populations. Since P2 is initially more compact and denser than P1, its binaries are affected by shorter relaxation timescales and higher encounter rates. This leads to more rapid segregation and enhanced disruption rates in the central regions of the cluster \citep{Bruce+2026}, producing the strong central peak and deeper deficit at intermediate radii observed in the profile of the total binary fraction. Concurrently, the more extended P1 population resides in regions characterized by lower local velocity dispersions and longer local relaxation timescales, allowing a larger fraction of wide P1 binaries to survive in the outskirts, preserving and sharpening the increase in the outer binary fraction. The presence of these dynamically distinct subsystems therefore strengthens the central peak, deepens the intermediate minimum, and maintains an increased fraction in the outer regions. The resulting bimodality is much more well defined and longer lived, persisting through the entirety of the simulation. In Appendix~\ref{sec:B1} we compare the binary fraction profile of this model with those of two different models, one starting with a different value of the P2 to the P1 initial half-mass radii and one with a smaller binary fraction: all models show a similar bimodal behavior.

\begin{figure}
    \centering
    \includegraphics[width=0.95\linewidth]{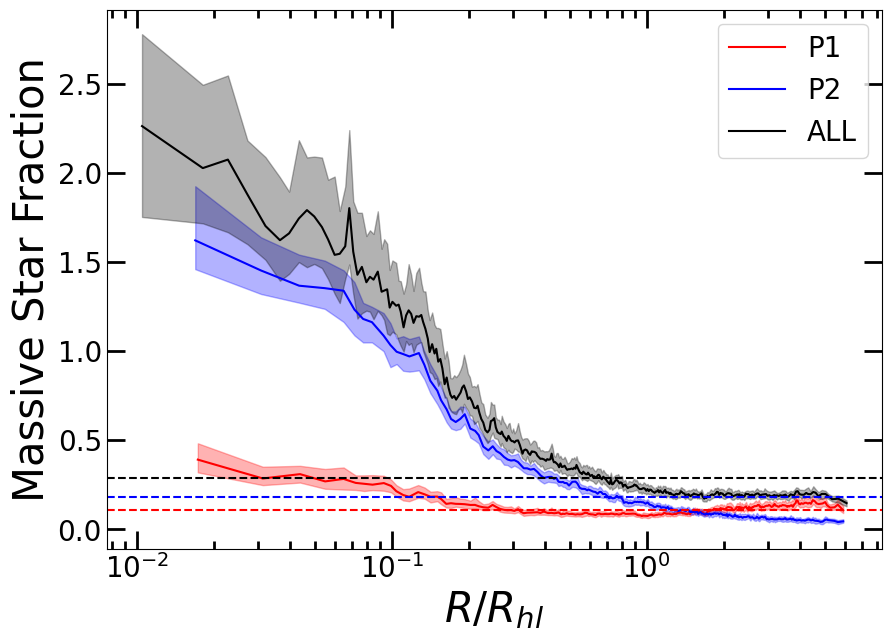}
    \caption{Radial profile of the fraction of massive stars (Equation 2) as a function of the projected distance from the cluster center normalized to the projected half-light radius for the MP simulation at 12 Gyr. We report the median of 100 random realizations of the 2D spatial projection, with the shaded regions representing the 25th and 75th percentiles. The horizontal dashed lines display the global values of the massive star fraction for P1 (red dashed line), P2 (blue dashed line) and P1+P2 (black dashed line). }
    \label{fig:fig 3}
\end{figure}

\begin{figure}[h!]
    \centering
    \includegraphics[width=0.95\linewidth, height=0.65\textheight]{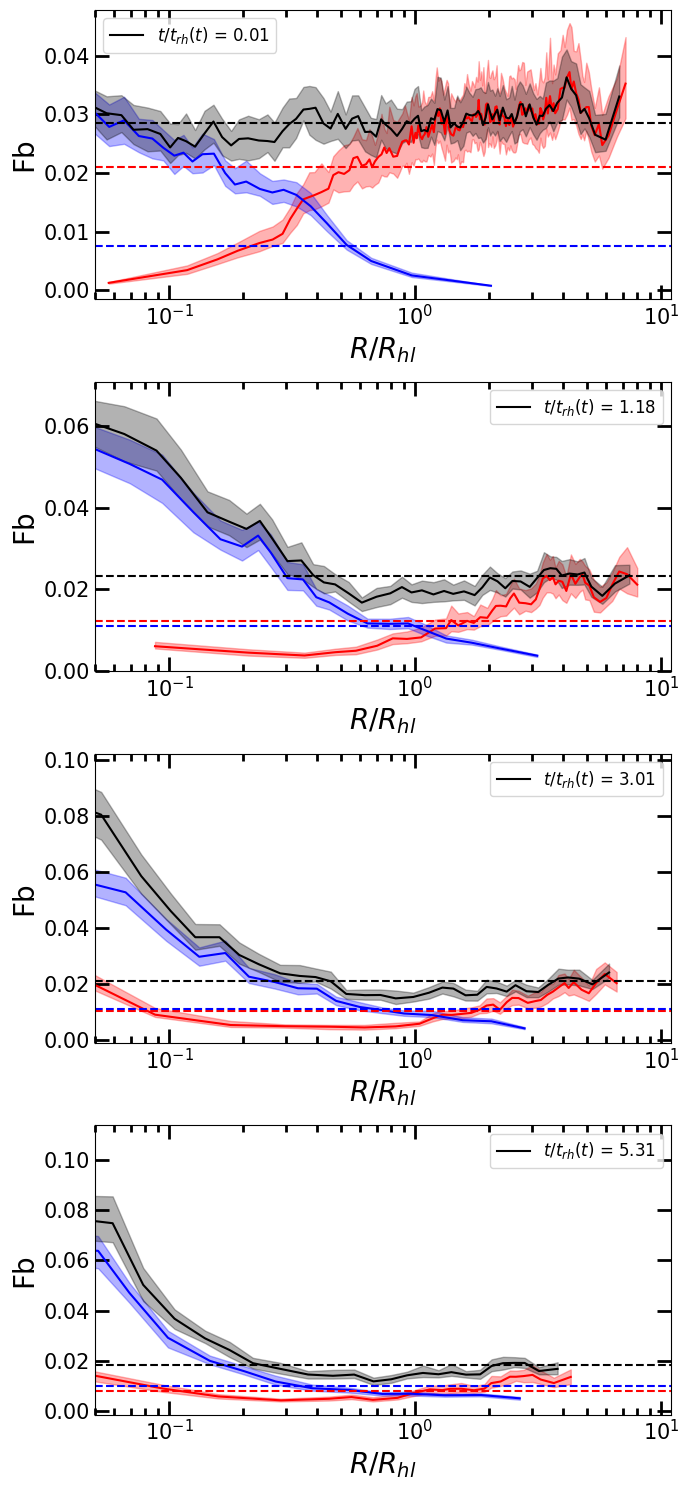}
    \caption{Radial profile of the fraction of compact binaries ($a<1$ AU) in P1 (red), P2 (blue), and both populations (black) as a function of the projected distance from the cluster center normalized to the projected half-light radius for the MP simulation at four different times. We report the median of 100 random realizations of the 2D spatial projection, with the shaded regions representing the 25th and 75th percentiles. The horizontal dashed lines display the global values of the binary fraction for P1 (red dashed line), P2 (blue dashed line) and P1+P2 (black dashed line). }
    \label{fig:fig 4}
\end{figure}

\begin{figure}[h!]
    \centering
    \includegraphics[width=0.95\linewidth, height=0.65\textheight]{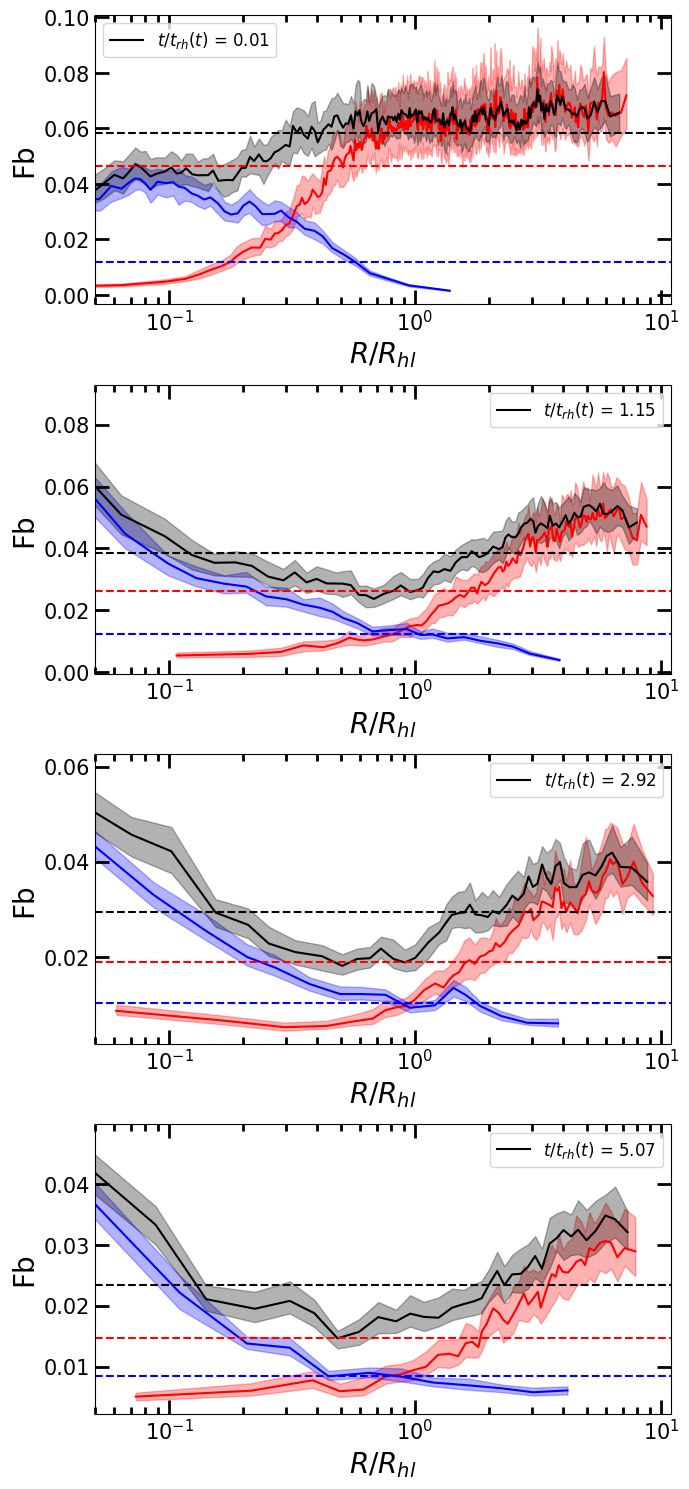}
    \caption{Radial profile of the fraction of binaries in P1 (red), P2 (blue), and both populations (black) as a function of the projected distance from the cluster center normalized to the projected half-light radius for the MPwide simulation at four different times. We report the median of 100 random realizations of the 2D spatial projection, with the shaded regions representing the 25th and 75th percentiles. The horizontal dashed lines display the global values of the binary fraction for P1 (red dashed line), P2 (blue dashed line) and P1+P2 (black dashed line). }
    \label{fig:fig 5}
\end{figure}

To better illustrate the role of the dynamical processes at play and disentangle the effects of segregation and binary disruption, we examined the radial distribution of single massive stars in the MP model. We focus on stars within 0.1 $M_\odot$ of the main sequence turnoff ($N^{High}$) at 12 Gyr rather than binary systems, and supplement the total number of objects considered using lower mass main sequence stars between 0.4 $M_\odot$ and 0.5 $M_\odot$ below the main sequence turnoff ($N^{Low}$). We then report the fraction of massive stars as

\begin{equation}
    Fraction  = \frac{N^{High}}{N^{High}+N^{Low}}.
\end{equation}

\noindent Replacing binaries with massive single stars allows us to trace the effects of mass segregation without the additional complexity of binary disruption or hardening. The resulting radial profile displayed in Figure \ref{fig:fig 3} is monotonic, where the fraction increases toward the center as higher mass stars segregate inward, and lower mass stars migrate outward.

To connect this investigation into mass segregation with binary stars, we repeated the analysis of the MP simulation using only the most compact binaries, with semimajor axes below 1 AU (Figure \ref{fig:fig 4}). These binaries are dynamically harder and are more likely to survive at almost all radii. They also interact with other objects in a more similar way to single stars of comparable combined mass (although some disruption or recoil may still occur during close binary-binary and binary-single encounters in the innermost regions). The resulting radial profile exhibits a strong central enhancement and a relatively flat outer region, closely resembling the profile of massive single stars. Notably, it does not show a pronounced bimodality, supporting that mass segregation alone can reproduce the central peak but is unable to generate a bimodality, and that, as discussed above, binary disruption is also required to form a significant minimum.

The MPwide simulation further illustrates the role of binary disruption. This model (displayed in Figure \ref{fig:fig 5}) contains a larger fraction of wide binaries with semimajor axes up to 1000 AU. The increased supply of wide binary systems results in an increased amount of binary disruption, strengthening the disruption driven deficit at small and intermediate radii, and producing a deeper minimum. The widest binaries can survive only in the outermost regions, whereas at smaller radii an increasing amount are soft and therefore efficiently destroyed. This model highlights the crucial role of binary disruption in shaping the radial structure of the binary fraction.

\begin{figure}
    \centering
    \includegraphics[width=0.95\linewidth, height=0.3\textheight]{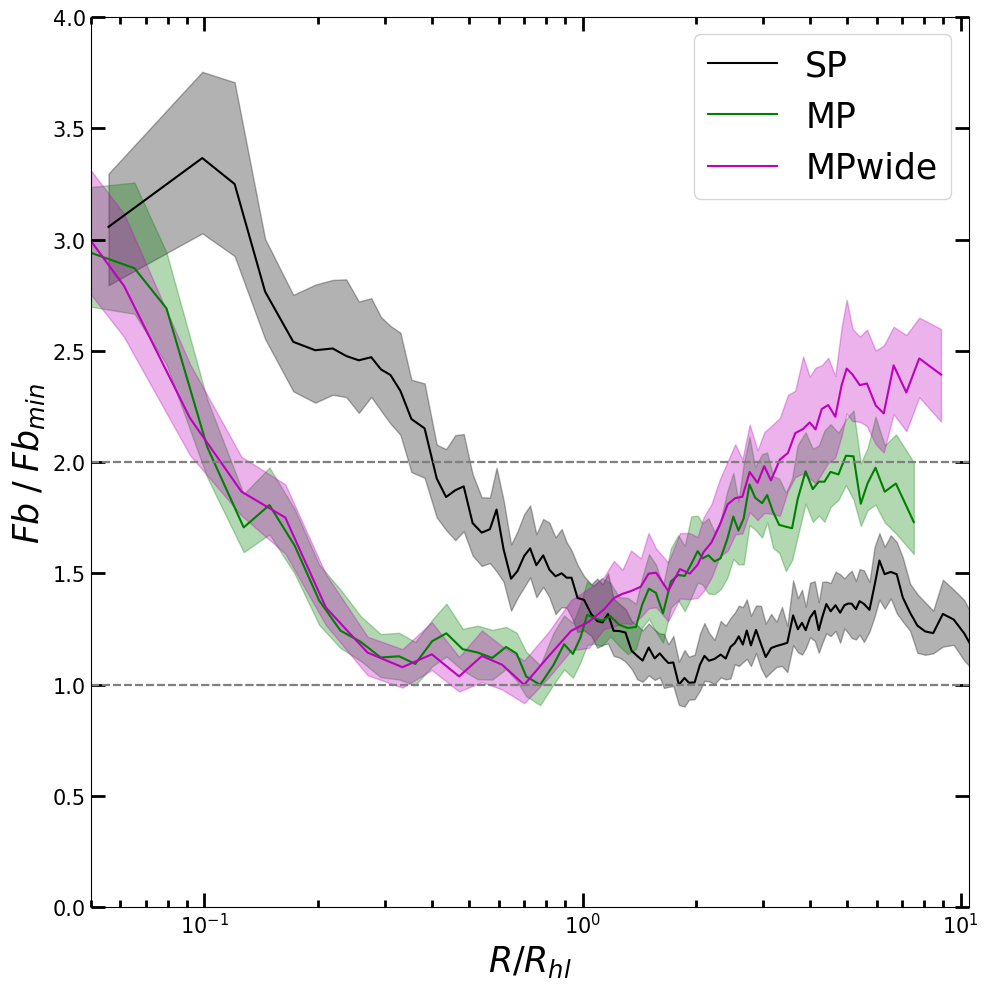}
    \caption{Radial profile of the fraction of binaries in the SP (black), MP (green), and MPwide (magenta) simulations as a function of the projected distance from the cluster center normalized to the projected half-light radius at a similar dynamical age ($t/t_{rh}(t)\sim2.5$). We report the median of 100 random realizations of the 2D spatial projection, with the shaded regions representing the 25th and 75th percentiles.}
    \label{fig:fig 6}
\end{figure}

Finally, to further illustrate the difference between the radial variation of the binary fraction in our models, Figure \ref{fig:fig 6} compares binary fraction profiles for the SP, MP, and MPwide simulations (taken at a dynamical age $t/t_{rh}(t)\sim\;2.5$) with each profile normalized to its minimum binary fraction. The SP model exhibits a monotonic radial trend with only a very mild outer increase as discussed in Sect. \ref{sec:3.1}, whereas both MP models retain a more pronounced bimodal structure. In the MP and MPwide simulations, the binary fraction increases by a factor of approximately 2 and 2.5 respectively in the outer regions relative to the minimum. Many of the clusters analyzed in paper I show a similar increase in the outer regions (see their Figure 5).

\subsection{Binary fraction minimum location}
\label{sec:3.3}

\begin{figure*}[h!]
    \centering
    \includegraphics[height=0.2\textheight,width=0.95\linewidth]{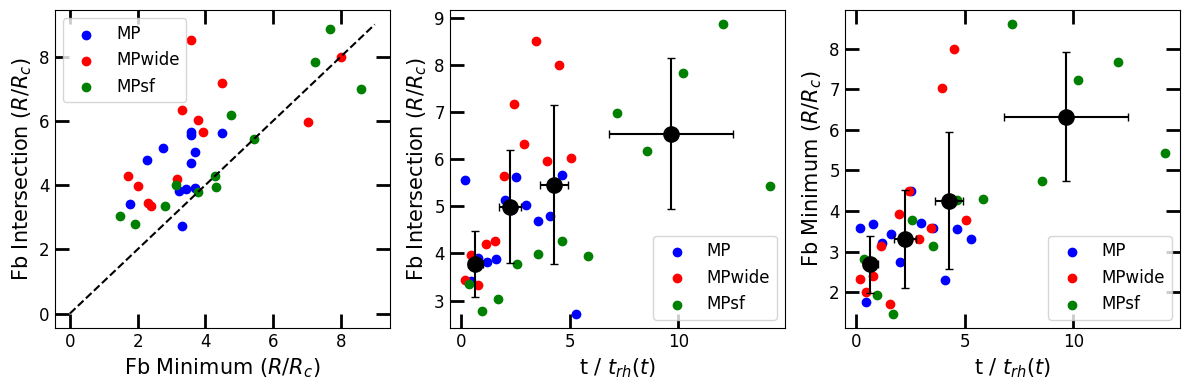}
    \caption{(Left) The correlation between the location of the minimum of the binary fraction radial profile, and the location of the intersection of the binary fractions within P1 and P2. (Middle) Time evolution of the intersection of the P1 and P2 binary fractions. (Right) Time evolution of the location of the binary fraction radial profile minimum. Location of the binary fraction minimum was evaluated visually each timestep. Black points show the mean locations of groups of 10 data points (6 data points in the final grouping) across all simulations, with error bars showing the standard deviation, and are included to guide the eye regarding the overall trend.}
    \label{fig:fig 7}
\end{figure*}

In this section we investigate the possible relationship between the shape of the bimodal profile and the dynamical age of the cluster. Specifically, following the observational analysis of paper I we analyze the time evolution of the minimum of the radial distribution of the binary fraction. However, in an attempt to be more robust and overcome the uncertainties and limitations in the determination of the binary fraction minimum, we also use the intersection of the radial binary fractions within the two populations as a proxy for the location of the minimum, as there is a correlation in their radial locations within the clusters. The left panel of Figure \ref{fig:fig 7} shows this correlation between the radial location of the minimum in the total binary fraction and the radius at which the P1 and P2 binary fractions intersect. We point out that in our analysis the location of the binary minimum is identified visually, and contains a large degree of uncertainty. However, these results are reported just to provide a general indication of the dynamical evolution of this feature and its relationship with the cluster dynamical age rather than specific locations.

The two additional panels of Figure \ref{fig:fig 7} show the time evolution of the radial positions of the minimum and the P1 and P2 binary fraction intersection. As the cluster evolves, both positions tend to move outward, reflecting the longer relaxation timescales at larger radii becoming increasingly relevant. Additionally, as both the inward segregation and disruption of wider binaries become more efficient at larger radii over time, there is a general shift of the region where the binary fraction is lowest. These radial positions are driven by the effects of the long-term dynamical evolution and provide a broad indication of the dynamical age of the cluster, although determining the location of these points may prove difficult in practice and lead to a large scatter in the observed trend.

\section{Conclusions}
\label{sec:4}

This paper provides insight into the dynamical processes responsible for shaping and preserving the bimodal radial profile of the binary fraction revealed by the observational study presented in paper I. By following the long term evolution of both single-population (SP) and multiple-population (MP) globular clusters, we investigated how dynamical processes such as binary disruption and mass segregation generate the bimodal distribution, and how the distinct environments inhabited by P1 and P2 binaries strengthen and prolong this feature.

Figure \ref{fig:fig 1} showcases the ability of SP clusters to produce a bimodal profile only as a transient feature. At early times, the efficient disruption of soft binaries in the dense core, combined with the survival of wide binaries in the outskirts and the inward segregation of compact binaries, can briefly create a weak bimodality. However, this profile is quickly replaced by a monotonic profile as binaries migrating inward from the outer regions refill the deficit at intermediate radii. A mild secondary bimodality can reappear 1-2 Gyr after core collapse due to recoil driven scattering from interactions in the core, but this requires the cluster to evolve past core collapse and the produced profile is still weaker and more centrally concentrated than the observed one. This might explain the bimodal radial distribution of the binary fraction observed by \citet{Cadelano+2026PREP} in the post-core collapse cluster M30.

Conversely, the presence of multiple populations in stellar clusters turns out to be a key ingredient in producing the long-lived bimodal binary distributions. In fact, the combination of binary disruption and segregation acting in a cluster with a structure determined by the presence of a dense P2 subsystem embedded in a more extended P1 system produces a clear and long-lived bimodality (Fig. \ref{fig:fig 2}). We show that although the initially more compact P2 population results in the more efficient segregation of binaries toward the central regions enhancing the central peak, mass segregation alone can not account for the full bimodality (see Fig. \ref{fig:fig 3} and Fig. \ref{fig:fig 4}). The key role of disruption was further illustrated with a simulation starting with a larger fraction of wider binaries where we find the more extended P1 population preserves wide binaries in the outer regions, maintaining a more pronounced minimum and increasing the outer binary fraction (see Fig. \ref{fig:fig 5} and Fig. \ref{fig:fig 6}).

Finally, we show that the effects of long-term dynamical evolution drive a broad trend between the radial location of the binary fraction minimum, the radius at which the P1 and P2 binary fractions intersect, and the dynamical age of the cluster (see Fig.\ref{fig:fig 7}). As the clusters evolve, the location of the minimum and the point of intersection tend to move outward, reflecting the dynamical processes becoming gradually more relevant at larger clustercentric distances over time.

The results presented in this paper showcase how binary stars are powerful tracers of globular cluster dynamics and of the formation and evolution of multiple stellar populations. Their radial distributions retain clear imprints of the cluster's initial configuration, as well as the dynamical processes affecting their long term evolution. The observed bimodality in the binary fraction naturally arises from the combination of mass segregation, binary disruption, and the distinct dynamical environments associated with multiple stellar populations.

\section{Acknowledgments}

EV acknowledges support from NSF grant AST-2009193. This
research was supported in part by Lilly Endowment, Inc., 570
through its support for the Indiana University Pervasive Technology Institute. EV acknowledges support from the John and A-Lan Reynolds Faculty Research Fund. ED acknowledges financial support from the INAF Data analysis Research Grant (PI E. Dalessandro) of the “Bando Astrofisica Fondamentale 2024”.

\bibliographystyle{aa}
\bibliography{ref}

@article{Bruce+2026,
	author = {{Bruce}, J and {Vesperini, E.} and {Askar, A.} and {Bortolan, E.} and {Giersz, M.} and {Hong, J.} and {Hypki, A.} and {Milone, A. P.}},
	title = {Exploring the dynamical evolution of binary stars in multiple-population globular clusters},
	DOI= "10.1051/0004-6361/202557826",
	url= "https://doi.org/10.1051/0004-6361/202557826",
	journal = {A\&A},
	year = 2026,
	volume = 707,
	pages = "A284",
}

@ARTICLE{Askar+2018,
       author = {{Askar}, Abbas and {Arca Sedda}, Manuel and {Giersz}, Mirek},
        title = "{MOCCA-SURVEY Database I: Galactic globular clusters harbouring a black hole subsystem}",
      journal = {\mnras},
         year = 2018,
        month = aug,
       volume = {478},
       number = {2},
        pages = {1844-1854},
          doi = {10.1093/mnras/sty1186},
archivePrefix = {arXiv},
       eprint = {1802.05284},
 primaryClass = {astro-ph.GA},
       adsurl = {https://ui.adsabs.harvard.edu/abs/2018MNRAS.478.1844A}
}

@ARTICLE{Kremer+2018,
       author = {{Kremer}, Kyle and {Ye}, Claire S. and {Chatterjee}, Sourav and {Rodriguez}, Carl L. and {Rasio}, Frederic A.},
        title = "{How Black Holes Shape Globular Clusters: Modeling NGC 3201}",
      journal = {\apjl},
         year = 2018,
        month = mar,
       volume = {855},
       number = {2},
          eid = {L15},
        pages = {L15},
          doi = {10.3847/2041-8213/aab26c},
archivePrefix = {arXiv},
       eprint = {1802.09553},
 primaryClass = {astro-ph.HE},
       adsurl = {https://ui.adsabs.harvard.edu/abs/2018ApJ...855L..15K}
}

@ARTICLE{Aros+2021,
       author = {{Aros}, Francisco I. and {Sippel}, Anna C. and {Mastrobuono-Battisti}, Alessandra and {Bianchini}, Paolo and {Askar}, Abbas and {van de Ven}, Glenn},
        title = "{Using binaries in globular clusters to catch sight of intermediate-mass black holes}",
      journal = {\mnras},
         year = 2021,
        month = dec,
       volume = {508},
       number = {3},
        pages = {4385-4398},
          doi = {10.1093/mnras/stab2872},
archivePrefix = {arXiv},
       eprint = {2110.00590},
 primaryClass = {astro-ph.GA},
       adsurl = {https://ui.adsabs.harvard.edu/abs/2021MNRAS.508.4385A}
}

@ARTICLE{Aros+2023,
       author = {{Aros}, Francisco I. and {Vesperini}, Enrico},
        title = "{Effects of massive central objects on the degree of energy equipartition of globular clusters}",
      journal = {\mnras},
         year = 2023,
        month = oct,
       volume = {525},
       number = {2},
        pages = {3136-3148},
          doi = {10.1093/mnras/stad2429},
archivePrefix = {arXiv},
       eprint = {2308.03845},
 primaryClass = {astro-ph.GA},
       adsurl = {https://ui.adsabs.harvard.edu/abs/2023MNRAS.525.3136A}
}

@ARTICLE{dellacroce+2024,
       author = {{Della Croce}, A. and {Aros}, F.~I. and {Vesperini}, E. and {Dalessandro}, E. and {Lanzoni}, B. and {Ferraro}, F.~R. and {Bhat}, B.},
        title = "{Inference of black-hole mass fraction in Galactic globular clusters: A multi-dimensional approach to break the initial-condition degeneracies}",
      journal = {\aap},
         year = 2024,
        month = oct,
       volume = {690},
          eid = {A179},
        pages = {A179},
          doi = {10.1051/0004-6361/202450954},
archivePrefix = {arXiv},
       eprint = {2409.01400},
 primaryClass = {astro-ph.GA},
       adsurl = {https://ui.adsabs.harvard.edu/abs/2024A&A...690A.179D}
}

@ARTICLE{Breen+2013,
       author = {{Breen}, Philip G. and {Heggie}, Douglas C.},
        title = "{On black hole subsystems in idealized nuclear star clusters}",
      journal = {\mnras},
         year = 2013,
        month = nov,
       volume = {436},
       number = {1},
        pages = {584-589},
          doi = {10.1093/mnras/stt1599},
archivePrefix = {arXiv},
       eprint = {1308.4641},
 primaryClass = {astro-ph.GA},
       adsurl = {https://ui.adsabs.harvard.edu/abs/2013MNRAS.436..584B}
}

@ARTICLE{Cadelano+2026PREP,
       author = {{Cadelano}, M and {Dalessandro}, E and {Bruce}, J and {Vesperini}, E and {Ferraro}, F.R. and {Lanzoni}, B and {Beccari}, G and {Giusti}, C and {Cusano}, F and {Paris}, D},
        title = "{Binary Stars as Dynamical Tracers in Globular Clusters .I. First Observations of Bimodal Spatial Distributions}",
      journal = {A\&A, in press},
         year = 2026,
        month = Submitted,
          eid = {},
        pages = {},
          doi = {},
archivePrefix = {arXiv},
       eprint = {},
 primaryClass = {},
       adsurl = {}
}

@ARTICLE{Geller+2013,
       author = {{Geller}, Aaron M. and {de Grijs}, Richard and {Li}, Chengyuan and {Hurley}, Jarrod R.},
        title = "{Consequences of Dynamical Disruption and Mass Segregation for the Binary Frequencies of Star Clusters}",
      journal = {\apj},
         year = 2013,
        month = dec,
       volume = {779},
       number = {1},
          eid = {30},
        pages = {30},
          doi = {10.1088/0004-637X/779/1/30},
archivePrefix = {arXiv},
       eprint = {1310.1085},
 primaryClass = {astro-ph.SR},
       adsurl = {https://ui.adsabs.harvard.edu/abs/2013ApJ...779...30G}
}

@ARTICLE{Carretta+2009a,
       author = {{Carretta}, E. and {Bragaglia}, A. and {Gratton}, R.~G. and {Lucatello}, S. and {Catanzaro}, G. and {Leone}, F. and {Bellazzini}, M. and {Claudi}, R. and {D'Orazi}, V. and {Momany}, Y. and {Ortolani}, S. and {Pancino}, E. and {Piotto}, G. and {Recio-Blanco}, A. and {Sabbi}, E.},
        title = "{Na-O anticorrelation and HB. VII. The chemical composition of first and second-generation stars in 15 globular clusters from GIRAFFE spectra}",
      journal = {\aap},
         year = 2009,
        month = oct,
       volume = {505},
       number = {1},
        pages = {117-138},
          doi = {10.1051/0004-6361/200912096},
archivePrefix = {arXiv},
       eprint = {0909.2938},
 primaryClass = {astro-ph.GA},
       adsurl = {https://ui.adsabs.harvard.edu/abs/2009A&A...505..117C}
}

@ARTICLE{Carretta2019,
       author = {{Carretta}, Eugenio},
        title = "{Empirical estimates of the Na-O anti-correlation in 95 Galactic globular clusters}",
      journal = {\aap},
         year = 2019,
        month = apr,
       volume = {624},
          eid = {A24},
        pages = {A24},
          doi = {10.1051/0004-6361/201935110},
archivePrefix = {arXiv},
       eprint = {1903.04494},
 primaryClass = {astro-ph.SR},
       adsurl = {https://ui.adsabs.harvard.edu/abs/2019A&A...624A..24C}
}

@ARTICLE{Marino+2019,
       author = {{Marino}, A.~F. and {Milone}, A.~P. and {Renzini}, A. and {D'Antona}, F. and {Anderson}, J. and {Bedin}, L.~R. and {Bellini}, A. and {Cordoni}, G. and {Lagioia}, E.~P. and {Piotto}, G. and {Tailo}, M.},
        title = "{The Hubble Space Telescope UV Legacy Survey of Galactic Globular Clusters - XIX. A chemical tagging of the multiple stellar populations over the chromosome maps}",
      journal = {\mnras},
         year = 2019,
        month = aug,
       volume = {487},
       number = {3},
        pages = {3815-3844},
          doi = {10.1093/mnras/stz1415},
archivePrefix = {arXiv},
       eprint = {1904.05180},
 primaryClass = {astro-ph.SR},
       adsurl = {https://ui.adsabs.harvard.edu/abs/2019MNRAS.487.3815M}
}

@ARTICLE{Kamann+2020,
       author = {{Kamann}, S. and {Giesers}, B. and {Bastian}, N. and {Brinchmann}, J. and {Dreizler}, S. and {G{\"o}ttgens}, F. and {Husser}, T.-O. and {Latour}, M. and {Weilbacher}, P.~M. and {Wisotzki}, L.},
        title = "{The binary content of multiple populations in NGC 3201}",
      journal = {\aap},
         year = 2020,
        month = mar,
       volume = {635},
          eid = {A65},
        pages = {A65},
          doi = {10.1051/0004-6361/201936843},
archivePrefix = {arXiv},
       eprint = {1912.01627},
 primaryClass = {astro-ph.SR},
       adsurl = {https://ui.adsabs.harvard.edu/abs/2020A&A...635A..65K}
}

@BOOK{HeggieHut2003,
       author = {{Heggie}, Douglas and {Hut}, Piet},
        title = "{The Gravitational Million-Body Problem: A Multidisciplinary Approach to Star Cluster Dynamics}",
         year = 2003,
       adsurl = {https://ui.adsabs.harvard.edu/abs/2003gmbp.book.....H}
}

@ARTICLE{Bortolan+2025,
       author = {{Bortolan}, E. and {Bruce}, J. and {Milone}, A.~P. and {Vesperini}, E. and {Dondoglio}, E. and {Legnardi}, M.~V. and {Muratore}, F. and {Ziliotto}, T. and {Cordoni}, G. and {Lagioia}, E.~P. and {Marino}, A.~F. and {Tailo}, M.},
        title = "{Exploring the formation environment and dynamics of multiple stellar populations in globular clusters through binary systems}",
      journal = {\aap},
         year = 2025,
        month = apr,
       volume = {696},
          eid = {A220},
        pages = {A220},
          doi = {10.1051/0004-6361/202452786},
archivePrefix = {arXiv},
       eprint = {2503.09708},
 primaryClass = {astro-ph.GA},
       adsurl = {https://ui.adsabs.harvard.edu/abs/2025A&A...696A.220B}
}

@ARTICLE{Hong+2015,
       author = {{Hong}, Jongsuk and {Vesperini}, Enrico and {Sollima}, Antonio and {McMillan}, Stephen L.~W. and {D'Antona}, Franca and {D'Ercole}, Annibale},
        title = "{Evolution of binary stars in multiple-population globular clusters}",
      journal = {\mnras},
         year = 2015,
        month = may,
       volume = {449},
       number = {1},
        pages = {629-638},
          doi = {10.1093/mnras/stv306},
archivePrefix = {arXiv},
       eprint = {1503.02087},
 primaryClass = {astro-ph.GA},
       adsurl = {https://ui.adsabs.harvard.edu/abs/2015MNRAS.449..629H}
}

@ARTICLE{Hong+2016,
       author = {{Hong}, Jongsuk and {Vesperini}, Enrico and {Sollima}, Antonio and {McMillan}, Stephen L.~W. and {D'Antona}, Franca and {D'Ercole}, Annibale},
        title = "{Evolution of binary stars in multiple-population globular clusters - II. Compact binaries}",
      journal = {\mnras},
         year = 2016,
        month = apr,
       volume = {457},
       number = {4},
        pages = {4507-4514},
          doi = {10.1093/mnras/stw262},
archivePrefix = {arXiv},
       eprint = {1604.01045},
 primaryClass = {astro-ph.GA},
       adsurl = {https://ui.adsabs.harvard.edu/abs/2016MNRAS.457.4507H}
}

@ARTICLE{Milone+2018b,
       author = {{Milone}, A.~P. and {Marino}, A.~F. and {Renzini}, A. and {D'Antona}, F. and {Anderson}, J. and {Barbuy}, B. and {Bedin}, L.~R. and {Bellini}, A. and {Brown}, T.~M. and {Cassisi}, S. and {Cordoni}, G. and {Lagioia}, E.~P. and {Nardiello}, D. and {Ortolani}, S. and {Piotto}, G. and {Sarajedini}, A. and {Tailo}, M. and {van der Marel}, R.~P. and {Vesperini}, E.},
        title = "{The Hubble Space Telescope UV legacy survey of galactic globular clusters - XVI. The helium abundance of multiple populations}",
      journal = {\mnras},
         year = 2018,
        month = dec,
       volume = {481},
       number = {4},
        pages = {5098-5122},
          doi = {10.1093/mnras/sty2573},
archivePrefix = {arXiv},
       eprint = {1809.05006},
 primaryClass = {astro-ph.SR},
       adsurl = {https://ui.adsabs.harvard.edu/abs/2018MNRAS.481.5098M}
}

@ARTICLE{Milone+2018a,
       author = {{Milone}, A.~P. and {Marino}, A.~F. and {Mastrobuono-Battisti}, A. and {Lagioia}, E.~P.},
        title = "{Gaia unveils the kinematics of multiple stellar populations in 47 Tucanae}",
      journal = {\mnras},
         year = 2018,
        month = oct,
       volume = {479},
       number = {4},
        pages = {5005-5011},
          doi = {10.1093/mnras/sty1873},
archivePrefix = {arXiv},
       eprint = {1807.03511},
 primaryClass = {astro-ph.SR},
       adsurl = {https://ui.adsabs.harvard.edu/abs/2018MNRAS.479.5005M}
}

@ARTICLE{Milone+2020,
       author = {{Milone}, A.~P. and {Vesperini}, E. and {Marino}, A.~F. and {Hong}, J. and {van der Marel}, R. and {Anderson}, J. and {Renzini}, A. and {Cordoni}, G. and {Bedin}, L.~R. and {Bellini}, A. and {Brown}, T.~M. and {D'Antona}, F. and {Lagioia}, E.~P. and {Libralato}, M. and {Nardiello}, D. and {Piotto}, G. and {Tailo}, M. and {Cool}, A. and {Salaris}, M. and {Sarajedini}, A.},
        title = "{The Hubble Space Telescope UV Legacy Survey of Galactic globular clusters - XXI. Binaries among multiple stellar populations}",
      journal = {\mnras},
         year = 2020,
        month = mar,
       volume = {492},
       number = {4},
        pages = {5457-5469},
          doi = {10.1093/mnras/stz3629},
archivePrefix = {arXiv},
       eprint = {2002.06479},
 primaryClass = {astro-ph.SR},
       adsurl = {https://ui.adsabs.harvard.edu/abs/2020MNRAS.492.5457M}
}

@ARTICLE{Gieles+2025,
       author = {{Gieles}, Mark and {Padoan}, Paolo and {Charbonnel}, Corinne and {Vink}, Jorick S. and {Ram{\'\i}rez-Galeano}, Laura},
        title = "{Globular cluster formation from inertial inflows: accreting extremely massive stars as the origin of abundance anomalies}",
      journal = {\mnras},
         year = 2025,
        month = nov,
       volume = {544},
       number = {1},
        pages = {483-512},
          doi = {10.1093/mnras/staf1314},
archivePrefix = {arXiv},
       eprint = {2501.12138},
 primaryClass = {astro-ph.GA},
       adsurl = {https://ui.adsabs.harvard.edu/abs/2025MNRAS.544..483G}
}

@ARTICLE{Bekki2010,
       author = {{Bekki}, Kenji},
        title = "{Rotation and Multiple Stellar Population in Globular Clusters}",
      journal = {\apjl},
         year = 2010,
        month = nov,
       volume = {724},
       number = {1},
        pages = {L99-L103},
          doi = {10.1088/2041-8205/724/1/L99},
archivePrefix = {arXiv},
       eprint = {1010.3841},
 primaryClass = {astro-ph.CO},
       adsurl = {https://ui.adsabs.harvard.edu/abs/2010ApJ...724L..99B}
}

@ARTICLE{Baumgardt+2023,
       author = {{Baumgardt}, H. and {H{\'e}nault-Brunet}, V. and {Dickson}, N. and {Sollima}, A.},
        title = "{Evidence for a bottom-light initial mass function in massive star clusters}",
      journal = {\mnras},
         year = 2023,
        month = may,
       volume = {521},
       number = {3},
        pages = {3991-4008},
          doi = {10.1093/mnras/stad631},
archivePrefix = {arXiv},
       eprint = {2303.01636},
 primaryClass = {astro-ph.GA},
       adsurl = {https://ui.adsabs.harvard.edu/abs/2023MNRAS.521.3991B}
}

@ARTICLE{Milone+2025,
       author = {{Milone}, A.~P. and {Marino}, A.~F. and {Bernizzoni}, M. and {Muratore}, F. and {Legnardi}, M.~V. and {Barbieri}, M. and {Bortolan}, E. and {Bouras}, A. and {Bruce}, J. and {Cordoni}, G. and {D'Antona}, F. and {Dell'Agli}, F. and {Dondoglio}, E. and {Grimaldi}, I.~M. and {Jang}, S. and {Lagioia}, E.~P. and {Lee}, J. -W. and {Lionetto}, S. and {Mohandasan}, A. and {Pang}, X. and {Pianta}, C. and {Posenato}, M. and {Renzini}, A. and {Tailo}, M. and {Ventura}, C. and {Ventura}, P. and {Vesperini}, E. and {Ziliotto}, T.},
        title = "{A JWST project on 47 Tucanae: Binaries among multiple populations}",
      journal = {\aap},
         year = 2025,
        month = jun,
       volume = {698},
          eid = {A247},
        pages = {A247},
          doi = {10.1051/0004-6361/202452136},
archivePrefix = {arXiv},
       eprint = {2503.19214},
 primaryClass = {astro-ph.SR},
       adsurl = {https://ui.adsabs.harvard.edu/abs/2025A&A...698A.247M}
}

@ARTICLE{Giersz1998,
       author = {{Giersz}, Mirek},
        title = "{Monte Carlo simulations of star clusters - I. First Results}",
      journal = {\mnras},
         year = 1998,
        month = aug,
       volume = {298},
       number = {4},
        pages = {1239-1248},
          doi = {10.1046/j.1365-8711.1998.01734.x},
archivePrefix = {arXiv},
       eprint = {astro-ph/9804127},
 primaryClass = {astro-ph},
       adsurl = {https://ui.adsabs.harvard.edu/abs/1998MNRAS.298.1239G}
}

@ARTICLE{Hypki+2013,
       author = {{Hypki}, Arkadiusz and {Giersz}, Mirek},
        title = "{MOCCA code for star cluster simulations - I. Blue stragglers, first results}",
      journal = {\mnras},
         year = 2013,
        month = feb,
       volume = {429},
       number = {2},
        pages = {1221-1243},
          doi = {10.1093/mnras/sts415},
archivePrefix = {arXiv},
       eprint = {1207.6700},
 primaryClass = {astro-ph.GA},
       adsurl = {https://ui.adsabs.harvard.edu/abs/2013MNRAS.429.1221H}
}

@article{Dercole+2008,
    author = {D'Ercole, Annibale and Vesperini, Enrico and D'Antona, Francesca and McMillan, Stephen L. W. and Recchi, Simone},
    title = {Formation and dynamical evolution of multiple stellar generations in globular clusters},
    journal = {MNRAS},
    volume = {391},
    number = {2},
    pages = {825-843},
    year = {2008},
    month = {11},
    issn = {0035-8711},
    doi = {10.1111/j.1365-2966.2008.13915.x},
    url = {https://doi.org/10.1111/j.1365-2966.2008.13915.x},
    eprint = {https://academic.oup.com/mnras/article-pdf/391/2/825/5773449/mnras0391-0825.pdf},
}

@ARTICLE{Hypki+2022,
       author = {{Hypki}, Arkadiusz and {Giersz}, Mirek and {Hong}, Jongsuk and {Leveque}, Agostino and {Askar}, Abbas and {Belloni}, Diogo and {Otulakowska-Hypka}, Magdalena},
        title = "{MOCCA: dynamics and evolution of single and binary stars of multiple stellar populations in tidally filling and underfilling globular star clusters}",
      journal = {\mnras},
         year = 2022,
        month = dec,
       volume = {517},
       number = {4},
        pages = {4768-4787},
          doi = {10.1093/mnras/stac2815},
archivePrefix = {arXiv},
       eprint = {2205.05397},
 primaryClass = {astro-ph.GA},
       adsurl = {https://ui.adsabs.harvard.edu/abs/2022MNRAS.517.4768H}
}

@article{Calura+2019,
    author = {Calura, F and D’Ercole, A and Vesperini, E and Vanzella, E and Sollima, A},
    title = {Formation of second-generation stars in globular clusters},
    journal = {MNRAS},
    volume = {489},
    number = {3},
    pages = {3269-3284},
    year = {2019},
    month = {07},
    issn = {0035-8711},
    doi = {10.1093/mnras/stz2055},
    url = {https://doi.org/10.1093/mnras/stz2055},
    eprint = {https://academic.oup.com/mnras/article-pdf/489/3/3269/30001914/stz2055.pdf},
}

@ARTICLE{Cordoni+2025binary,
       author = {{Cordoni}, Giacomo and {Casagrande}, Luca and {Jerjen}, Helmut},
        title = "{Rubin Data Preview 1: Extending the view of unresolved binary stars in 47 Tucanae}",
      journal = {\pasa},
         year = 2025,
        month = oct,
       volume = {42},
          eid = {e127},
        pages = {e127},
          doi = {10.1017/pasa.2025.10089},
archivePrefix = {arXiv},
       eprint = {2509.04054},
 primaryClass = {astro-ph.SR},
       adsurl = {https://ui.adsabs.harvard.edu/abs/2025PASA...42..127C}
}

@ARTICLE{Cadelano+2024,
       author = {{Cadelano}, Mario and {Dalessandro}, Emanuele and {Vesperini}, Enrico},
        title = "{The structural properties of multiple populations in globular clusters: The instructive case of NGC 3201}",
      journal = {\aap},
         year = 2024,
        month = may,
       volume = {685},
          eid = {A158},
        pages = {A158},
          doi = {10.1051/0004-6361/202349021},
archivePrefix = {arXiv},
       eprint = {2402.09514},
 primaryClass = {astro-ph.GA},
       adsurl = {https://ui.adsabs.harvard.edu/abs/2024A&A...685A.158C}
}

@ARTICLE{Cadelano+2020,
       author = {{Cadelano}, M. and {Dalessandro}, E. and {Webb}, J.~J. and {Vesperini}, E. and {Lattanzio}, D. and {Beccari}, G. and {Gomez}, M. and {Monaco}, L.},
        title = "{Radial variation of the stellar mass functions in the globular clusters M15 and M30: clues of a non-standard IMF?}",
      journal = {\mnras},
         year = 2020,
        month = dec,
       volume = {499},
       number = {2},
        pages = {2390-2400},
          doi = {10.1093/mnras/staa2759},
archivePrefix = {arXiv},
       eprint = {2009.02333},
 primaryClass = {astro-ph.GA},
       adsurl = {https://ui.adsabs.harvard.edu/abs/2020MNRAS.499.2390C}
}

@ARTICLE{Martens+2023,
       author = {{Martens}, Sven and {Kamann}, Sebastian and {Dreizler}, Stefan and {G{\"o}ttgens}, Fabian and {Husser}, Tim-Oliver and {Latour}, Marilyn and {Balakina}, Elena and {Krajnovi{\'c}}, Davor and {Pechetti}, Renuka and {Weilbacher}, Peter M.},
        title = "{Kinematic differences between multiple populations in Galactic globular clusters}",
      journal = {\aap},
         year = 2023,
        month = mar,
       volume = {671},
          eid = {A106},
        pages = {A106},
          doi = {10.1051/0004-6361/202244787},
archivePrefix = {arXiv},
       eprint = {2301.08675},
 primaryClass = {astro-ph.GA},
       adsurl = {https://ui.adsabs.harvard.edu/abs/2023A&A...671A.106M}
}

@ARTICLE{Cordoni+2025dynamics,
       author = {{Cordoni}, G. and {Casagrande}, L. and {Milone}, A.~P. and {Dondoglio}, E. and {Mastrobuono-Battisti}, A. and {Jang}, S. and {Marino}, A.~F. and {Lagioia}, E.~P. and {Legnardi}, M.~V. and {Ziliotto}, T. and {Muratore}, F. and {Mehta}, V. and {Lacchin}, E. and {Tailo}, M.},
        title = "{Internal dynamics of multiple populations in 28 Galactic globular clusters: a wide-field study with Gaia and the Hubble Space Telescope}",
      journal = {\mnras},
         year = 2025,
        month = mar,
       volume = {537},
       number = {3},
        pages = {2342-2361},
          doi = {10.1093/mnras/staf102},
archivePrefix = {arXiv},
       eprint = {2409.02330},
 primaryClass = {astro-ph.GA},
       adsurl = {https://ui.adsabs.harvard.edu/abs/2025MNRAS.537.2342C}
}

@ARTICLE{King1966,
       author = {{King}, Ivan R.},
        title = "{The structure of star clusters. III. Some simple dynamical models}",
      journal = {\aj},
         year = 1966,
        month = feb,
       volume = {71},
        pages = {64},
          doi = {10.1086/109857},
       adsurl = {https://ui.adsabs.harvard.edu/abs/1966AJ.....71...64K}
}

@ARTICLE{Kroupa2001,
       author = {{Kroupa}, Pavel},
        title = "{On the variation of the initial mass function}",
      journal = {\mnras},
         year = 2001,
        month = apr,
       volume = {322},
       number = {2},
        pages = {231-246},
          doi = {10.1046/j.1365-8711.2001.04022.x},
archivePrefix = {arXiv},
       eprint = {astro-ph/0009005},
 primaryClass = {astro-ph},
       adsurl = {https://ui.adsabs.harvard.edu/abs/2001MNRAS.322..231K}
}

@ARTICLE{Milone+2022,
       author = {{Milone}, Antonino P. and {Marino}, Anna F.},
        title = "{Multiple Populations in Star Clusters}",
      journal = {Universe},
         year = 2022,
        month = jun,
       volume = {8},
       number = {7},
          eid = {359},
        pages = {359},
          doi = {10.3390/universe8070359},
archivePrefix = {arXiv},
       eprint = {2206.10564},
 primaryClass = {astro-ph.GA},
       adsurl = {https://ui.adsabs.harvard.edu/abs/2022Univ....8..359M}
}

@ARTICLE{Gieles+2018,
       author = {{Gieles}, Mark and {Charbonnel}, Corinne and {Krause}, Martin G.~H. and {H{\'e}nault-Brunet}, Vincent and {Agertz}, Oscar and {Lamers}, Henny J.~G.~L.~M. and {Bastian}, Nathan and {Gualandris}, Alessia and {Zocchi}, Alice and {Petts}, James A.},
        title = "{Concurrent formation of supermassive stars and globular clusters: implications for early self-enrichment}",
      journal = {\mnras},
         year = 2018,
        month = aug,
       volume = {478},
       number = {2},
        pages = {2461-2479},
          doi = {10.1093/mnras/sty1059},
archivePrefix = {arXiv},
       eprint = {1804.04682},
 primaryClass = {astro-ph.GA},
       adsurl = {https://ui.adsabs.harvard.edu/abs/2018MNRAS.478.2461G}
}

@ARTICLE{Bastian+2013,
       author = {{Bastian}, N. and {Lamers}, H.~J.~G.~L.~M. and {de Mink}, S.~E. and {Longmore}, S.~N. and {Goodwin}, S.~P. and {Gieles}, M.},
        title = "{Early disc accretion as the origin of abundance anomalies in globular clusters}",
      journal = {\mnras},
         year = 2013,
        month = dec,
       volume = {436},
       number = {3},
        pages = {2398-2411},
          doi = {10.1093/mnras/stt1745},
archivePrefix = {arXiv},
       eprint = {1309.3566},
 primaryClass = {astro-ph.GA},
       adsurl = {https://ui.adsabs.harvard.edu/abs/2013MNRAS.436.2398B}
}

@ARTICLE{Bellini+2009,
       author = {{Bellini}, A. and {Piotto}, G. and {Bedin}, L.~R. and {King}, I.~R. and {Anderson}, J. and {Milone}, A.~P. and {Momany}, Y.},
        title = "{Radial distribution of the multiple stellar populations in {\ensuremath{\omega}} Centauri}",
      journal = {\aap},
         year = 2009,
        month = dec,
       volume = {507},
       number = {3},
        pages = {1393-1408},
          doi = {10.1051/0004-6361/200912757},
archivePrefix = {arXiv},
       eprint = {0909.4785},
 primaryClass = {astro-ph.SR},
       adsurl = {https://ui.adsabs.harvard.edu/abs/2009A&A...507.1393B}
}

@ARTICLE{simioni+2016,
       author = {{Simioni}, M. and {Milone}, A.~P. and {Bedin}, L.~R. and {Aparicio}, A. and {Piotto}, G. and {Vesperini}, E. and {Hong}, J.},
        title = "{The Hubble Space Telescope UV Legacy Survey of Galactic globular clusters - X. The radial distribution of stellar populations in NGC 2808}",
      journal = {\mnras},
         year = 2016,
        month = nov,
       volume = {463},
       number = {1},
        pages = {449-458},
          doi = {10.1093/mnras/stw2003},
archivePrefix = {arXiv},
       eprint = {1608.03124},
 primaryClass = {astro-ph.SR},
       adsurl = {https://ui.adsabs.harvard.edu/abs/2016MNRAS.463..449S}
}

@ARTICLE{Dalessandro+2019,
       author = {{Dalessandro}, Emanuele and {Cadelano}, M. and {Vesperini}, E. and {Martocchia}, S. and {Ferraro}, F.~R. and {Lanzoni}, B. and {Bastian}, N. and {Hong}, J. and {Sanna}, N.},
        title = "{A Family Picture: Tracing the Dynamical Path of the Structural Properties of Multiple Populations in Globular Clusters}",
      journal = {\apjl},
         year = 2019,
        month = oct,
       volume = {884},
       number = {1},
          eid = {L24},
        pages = {L24},
          doi = {10.3847/2041-8213/ab45f7},
archivePrefix = {arXiv},
       eprint = {1910.00613},
 primaryClass = {astro-ph.SR},
       adsurl = {https://ui.adsabs.harvard.edu/abs/2019ApJ...884L..24D}
}

@ARTICLE{Dalessandro+2024,
       author = {{Dalessandro}, E. and {Cadelano}, M. and {Della Croce}, A. and {Aros}, F.~I. and {White}, E.~B. and {Vesperini}, E. and {Fanelli}, C. and {Ferraro}, F.~R. and {Lanzoni}, B. and {Leanza}, S. and {Origlia}, L.},
        title = "{A 3D view of multiple populations' kinematics in Galactic globular clusters}",
      journal = {\aap},
         year = 2024,
        month = nov,
       volume = {691},
          eid = {A94},
        pages = {A94},
          doi = {10.1051/0004-6361/202451054},
archivePrefix = {arXiv},
       eprint = {2409.03827},
 primaryClass = {astro-ph.GA},
       adsurl = {https://ui.adsabs.harvard.edu/abs/2024A&A...691A..94D}
}

@ARTICLE{Dalessandro+2021,
       author = {{Dalessandro}, Emanuele and {Raso}, Silvia and {Kamann}, Sebastian and {Bellazzini}, Michele and {Vesperini}, Enrico and {Bellini}, Andrea and {Beccari}, Giacomo},
        title = "{3D core kinematics of NGC 6362: central rotation in a dynamically evolved globular cluster}",
      journal = {\mnras},
         year = 2021,
        month = sep,
       volume = {506},
       number = {1},
        pages = {813-823},
          doi = {10.1093/mnras/stab1257},
archivePrefix = {arXiv},
       eprint = {2105.02246},
 primaryClass = {astro-ph.GA},
       adsurl = {https://ui.adsabs.harvard.edu/abs/2021MNRAS.506..813D}
}

@ARTICLE{Onorato+2023,
       author = {{Onorato}, Silvia and {Cadelano}, Mario and {Dalessandro}, Emanuele and {Vesperini}, Enrico and {Lanzoni}, Barbara and {Mucciarelli}, Alessio},
        title = "{The structural properties of multiple populations in the dynamically young globular cluster NGC 2419}",
      journal = {\aap},
         year = 2023,
        month = sep,
       volume = {677},
          eid = {A8},
        pages = {A8},
          doi = {10.1051/0004-6361/202346792},
archivePrefix = {arXiv},
       eprint = {2307.09508},
 primaryClass = {astro-ph.GA},
       adsurl = {https://ui.adsabs.harvard.edu/abs/2023A&A...677A...8O}
}

@ARTICLE{Beccari+2013,
       author = {{Beccari}, G. and {Bellazzini}, M. and {Lardo}, C. and {Bragaglia}, A. and {Carretta}, E. and {Dalessandro}, E. and {Mucciarelli}, A. and {Pancino}, E.},
        title = "{Evidence for multiple populations in the massive globular cluster NGC 2419 from deep uVI LBT photometry}",
      journal = {\mnras},
         year = 2013,
        month = may,
       volume = {431},
       number = {2},
        pages = {1995-2005},
          doi = {10.1093/mnras/stt316},
archivePrefix = {arXiv},
       eprint = {1302.4375},
 primaryClass = {astro-ph.SR},
       adsurl = {https://ui.adsabs.harvard.edu/abs/2013MNRAS.431.1995B}
}

@ARTICLE{Hong+2019,
       author = {{Hong}, Jongsuk and {Patel}, Saahil and {Vesperini}, Enrico and {Webb}, Jeremy J. and {Dalessandro}, Emanuele},
        title = "{Spatial mixing of binary stars in multiple-population globular clusters}",
      journal = {\mnras},
         year = 2019,
        month = feb,
       volume = {483},
       number = {2},
        pages = {2592-2599},
          doi = {10.1093/mnras/sty3308},
archivePrefix = {arXiv},
       eprint = {1812.01229},
 primaryClass = {astro-ph.GA},
       adsurl = {https://ui.adsabs.harvard.edu/abs/2019MNRAS.483.2592H}
}

@ARTICLE{Vesperini+2011,
       author = {{Vesperini}, Enrico and {McMillan}, Stephen L.~W. and {D'Antona}, Francesca and {D'Ercole}, Annibale},
        title = "{Binary star disruption in globular clusters with multiple stellar populations}",
      journal = {\mnras},
         year = 2011,
        month = sep,
       volume = {416},
       number = {1},
        pages = {355-360},
          doi = {10.1111/j.1365-2966.2011.19046.x},
archivePrefix = {arXiv},
       eprint = {1106.0756},
 primaryClass = {astro-ph.GA},
       adsurl = {https://ui.adsabs.harvard.edu/abs/2011MNRAS.416..355V}
}

@ARTICLE{Lucatello+2015,
       author = {{Lucatello}, S. and {Sollima}, A. and {Gratton}, R. and {Vesperini}, E. and {D'Orazi}, V. and {Carretta}, E. and {Bragaglia}, A.},
        title = "{The incidence of binaries in globular cluster stellar populations}",
      journal = {\aap},
         year = 2015,
        month = dec,
       volume = {584},
          eid = {A52},
        pages = {A52},
          doi = {10.1051/0004-6361/201526957},
archivePrefix = {arXiv},
       eprint = {1509.05014},
 primaryClass = {astro-ph.SR},
       adsurl = {https://ui.adsabs.harvard.edu/abs/2015A&A...584A..52L}
}

@ARTICLE{dalessandro+2016,
       author = {{Dalessandro}, E. and {Lapenna}, E. and {Mucciarelli}, A. and {Origlia}, L. and {Ferraro}, F.~R. and {Lanzoni}, B.},
        title = "{Multiple Populations in the Old and Massive Small Magellanic Cloud Globular Cluster NGC 121}",
      journal = {\apj},
         year = 2016,
        month = oct,
       volume = {829},
       number = {2},
          eid = {77},
        pages = {77},
          doi = {10.3847/0004-637X/829/2/77},
archivePrefix = {arXiv},
       eprint = {1607.05736},
 primaryClass = {astro-ph.SR},
       adsurl = {https://ui.adsabs.harvard.edu/abs/2016ApJ...829...77D}
}

@ARTICLE{Dalessandro+2018b,
       author = {{Dalessandro}, E. and {Mucciarelli}, A. and {Bellazzini}, M. and {Sollima}, A. and {Vesperini}, E. and {Hong}, J. and {H{\'e}nault-Brunet}, Vincent and {Ferraro}, F.~R. and {Ibata}, R. and {Lanzoni}, B. and {Massari}, D. and {Salaris}, M.},
        title = "{The Unexpected Kinematics of Multiple Populations in NGC 6362: Do Binaries Play a Role?}",
      journal = {\apj},
         year = 2018,
        month = sep,
       volume = {864},
       number = {1},
          eid = {33},
        pages = {33},
          doi = {10.3847/1538-4357/aad4b3},
archivePrefix = {arXiv},
       eprint = {1807.07918},
 primaryClass = {astro-ph.GA},
       adsurl = {https://ui.adsabs.harvard.edu/abs/2018ApJ...864...33D}
}

@ARTICLE{Dalessandro+2018a,
       author = {{Dalessandro}, E. and {Cadelano}, M. and {Vesperini}, E. and {Salaris}, M. and {Ferraro}, F.~R. and {Lanzoni}, B. and {Raso}, S. and {Hong}, J. and {Webb}, J.~J. and {Zocchi}, A.},
        title = "{The Peculiar Radial Distribution of Multiple Populations in the Massive Globular Cluster M80}",
      journal = {\apj},
         year = 2018,
        month = may,
       volume = {859},
       number = {1},
          eid = {15},
        pages = {15},
          doi = {10.3847/1538-4357/aabb56},
archivePrefix = {arXiv},
       eprint = {1804.03222},
 primaryClass = {astro-ph.SR},
       adsurl = {https://ui.adsabs.harvard.edu/abs/2018ApJ...859...15D}
}

@ARTICLE{Bellini+2015,
       author = {{Bellini}, A. and {Vesperini}, E. and {Piotto}, G. and {Milone}, A.~P. and {Hong}, J. and {Anderson}, J. and {van der Marel}, R.~P. and {Bedin}, L.~R. and {Cassisi}, S. and {D'Antona}, F. and {Marino}, A.~F. and {Renzini}, A.},
        title = "{The Hubble Space Telescope UV Legacy Survey of Galactic Globular Clusters: The Internal Kinematics of the Multiple Stellar Populations in NGC 2808}",
      journal = {\apjl},
         year = 2015,
        month = sep,
       volume = {810},
       number = {1},
          eid = {L13},
        pages = {L13},
          doi = {10.1088/2041-8205/810/1/L13},
archivePrefix = {arXiv},
       eprint = {1508.01804},
 primaryClass = {astro-ph.GA},
       adsurl = {https://ui.adsabs.harvard.edu/abs/2015ApJ...810L..13B}
}

\begin{appendix}
\section{Cumulative mass radial profile}
\label{sec:A1}

\begin{figure}[h!]
    \centering
    \includegraphics[width=0.95\linewidth]{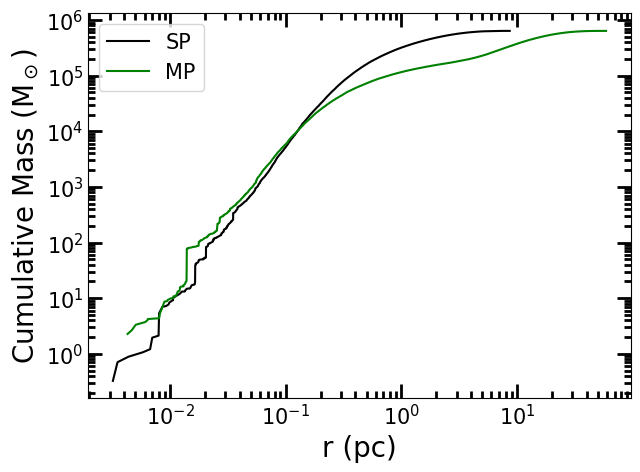}
    \caption{Initial cumulative radial mass profile for the SP (black) and MP (green) models as a function of clustercentric distance.}
    \label{fig:fig A1}
\end{figure}

In Figure~\ref{fig:fig A1} we show the initial cumulative radial mass profiles for the SP and MP models. The two models have similar mass distributions in the inner regions but the MP system has a multi-scale structure characterized by a centrally concentrated P2 population embedded in a more extended low-density P1 system.

\section{Bimodality for $r_{h, P2}/r_{h, P1}=0.1$ and $5\%$ binary fraction}
\label{sec:B1}

\begin{figure}[h!]
    \centering
    \includegraphics[width=0.95\linewidth]{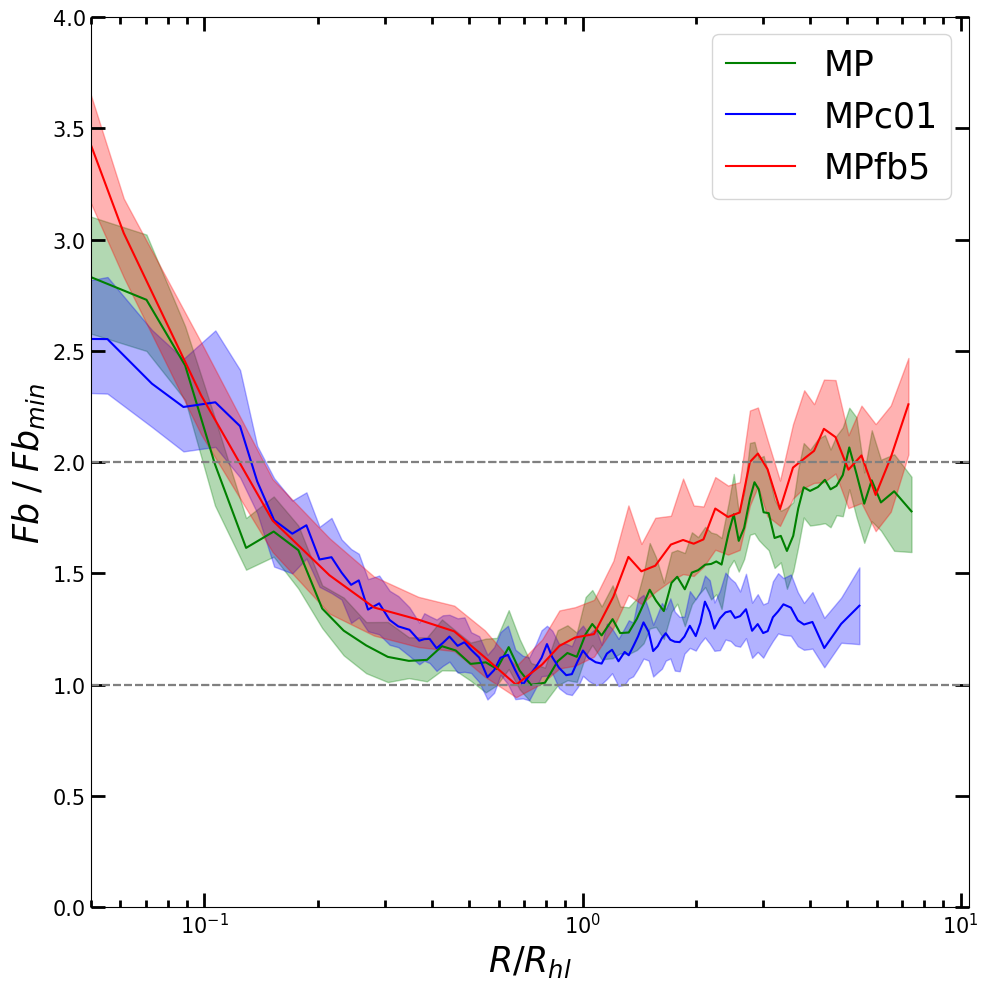}
    \caption{Radial profile of the fraction of binaries in the MP (green), MPc01 (blue), and MPfb5 (red) simulations as a function of the projected distance from the cluster center normalized to the projected half-light radius at a similar dynamical age \textbf{($t/t_{rh}(t)\sim2-2.5$)}. We report the median of 100 random realizations of the 2D spatial projection, with the shaded regions representing the 25th and 75th percentiles.}
    \label{fig:fig B1}
\end{figure}

In Figure~\ref{fig:fig B1} we show the results of two simulations with the same initial conditions adopted for the MP simulation (which has an initial binary fraction equal to $10\%$ and a P2 to P1 half-mass radii ratio equal to 0.05; see section~\ref{sec:2}) but different binary fraction or P2 to P1 half-mass radii ratio: one simulation starts with an initial ratio of the P2 to the P1 half-mass radii equal to 0.1 and $10\%$ binary fraction (MPc01) and one with P2 to P1 half-mass radii ratio equal to 0.05 and initial binary fraction equal to $5\%$ (MPfb5). In both cases the bimodality discussed in this paper is present with a minimum near the half-light radius. As discussed in the paper the multiscale structure of the multiple-population clusters is the key ingredient to produce the observed bimodality in the radial variation of the binary fraction. As the value of the half-mass radii ratio increases and the systems approach one with the structure of a single-population cluster the bimodality becomes milder. Additionally, a smaller initial binary fraction does not lead to significant difference in the strength of the bimodal behavior.

\end{appendix}

\end{document}